\documentclass[12pt]{iopart}

\usepackage{graphicx}

\begin{document}

\title{Quantum state transfer in arrays of flux qubits}

\author{A. Lyakhov and C. Bruder}

\address{Department of Physics and Astronomy, University of Basel,
 Klingelbergstr. 82, 4056 Basel, Switzerland}

\begin{abstract}
In this work, we describe a possible experimental realization of Bose's idea to
use spin chains for short distance quantum communication [S. Bose, {\it Phys.
Rev. Lett.} {\bf 91} 207901]. Josephson arrays have been proposed and analyzed
as transmission channels for systems of superconducting charge qubits. Here, we
consider a chain of persistent current qubits, that is appropriate for state
transfer with high fidelity in systems containing flux qubits. We calculate the
fidelity of state transfer for this system. In general, the Hamiltonian of this
system is not of XXZ-type, and we analyze the magnitude and the effect of the
terms that do not conserve the z-component of the total spin. 
\end{abstract}



\maketitle

\section{Introduction}

Recently, the idea to use quantum spin chains for short-distance quantum
communication was put forward by Bose \cite{bose03}. He showed that an array of
spins (or spin-like two level systems) with isotropic Heisenberg interaction is
suitable for quantum state transfer. The advantage of spin chains as
transmission lines is the fact, that they do not need to have controllable
couplings between the qubits or complicated gating schemes to achieve high
transfer fidelity. An initial state is prepared at one end of the chain at time
$t=0$, and after a certain time $t_{1}$ is measured at the other. The fidelity
of quantum communication averaged over all pure input states on the Bloch
sphere is taken as a measure of the transmission quality.

 Bose showed that for short chains (number of spins $\simeq$ 100) the average
fidelity is quite high, greater than 2/3, which is the highest fidelity of
transmission through a classical channel \cite{Horodecki}. In a homogeneous
chain, i.e. if all coupling constants are the same, the information about the
input state is dispersed between the spins at all times $t>0$. Therefore the
fidelity is always less then unity.

Some methods were proposed to achieve perfect state transfer with
fidelity one.  A special form of the Hamiltonian with spatially
varying coupling constants between the qubits allows to avoid
dispersion \cite{christandl,Paternostro}. Another method is to form
Gaussian wave packets (with low dispersion) by encoding the
information using multiple spins \cite{Osborne}.  Also the combination
of two spin chains \cite{bose04} can be used to achieve perfect state
transfer. This method has the advantage that it can be implemented
using almost any two spin chains and is stable to fluctuations of the
chain parameters \cite{bose05}. However, the time after which perfect
state transfer is achieved grows if the individual fidelities of the
chains decrease. It is therefore advantageous to have single chains
with high fidelity to implement this improved method.

Quantum state transfer can be implemented using any type of two-level systems.
However, it is preferable to use a technology that is adapted to the quantum
information hardware that is supposed to be coupled by the transmission line.
One of the most promising architectures of quantum computing devices are
superconducting circuits, for example charge, flux and charge-flux qubits. In
recent years these were intensively studied both theoretically and
experimentally.

 A possible realization of an effective transmission line for charge qubits was
described in \cite{romito}. There, the fidelity of state transfer through
Josephson junction arrays and the influence of static disorder and dynamical
noise was analyzed.

\section{Arrays of persistent-current qubits}

\begin{figure}
\begin{center}
 \includegraphics[width=0.6\textwidth]{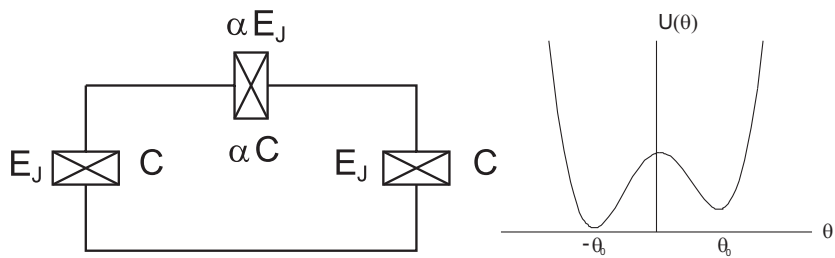}
\end{center}
\caption{Persistent-current qubit and Josephson energy as a function of the
relative phase of left and right Josephson junctions.}
\label{FluxQubit}
\end{figure}

In this work we consider a line of persistent-current qubits \cite{levitov}. We
will show that it is appropriate for state transfer with high fidelity in
systems containing flux qubits. A persistent current qubit \cite{mooij99} is a
superconducting loop with three Josephson junctions, see Fig. \ref{FluxQubit}.
We assume that the left and right Josephson junctions have capacitance $C$ and
Josephson energy $E_{J}$, the central junction is characterized by a
capacitance $\alpha C$ and Josephson energy $\alpha E_{J}$ with $\alpha <1$.
The gate capacitances (not shown in the figure) are equal $\gamma C$. The
Hamiltonian of the qubit

\begin{equation}
H_{0}=-\Delta_{0}\sigma^{x}-B\sigma^{z}
\end{equation}

is the same as that of a spin-$\frac{1}{2}$ particle in a magnetic field. The
eigenstates $|0\rangle\equiv|\downarrow\rangle$ and
$|1\rangle\equiv|\uparrow\rangle$ of $\sigma^{z}$ correspond to clockwise and
counterclockwise currents. The coefficient $\Delta_{0}$ is a tunneling
amplitude between these states and $B$ depends on the flux through the qubit
$\Phi$ and the modulus of the circulating current $I_{p}$

\begin{equation}
B=I_{p}(\Phi)\left(\Phi-\frac{1}{2}\Phi_{0}\right)\; ,
\label{bfield}
\end{equation}

here,  $\Phi_{0}=h/(2e)$ is the flux quantum.
The circulating current $I_{p}$ depends on
the magnetic frustration, i.e. the amount of external magnetic flux in the loop
in units of the flux quantum \cite{orlando99}. The effective magnetic field $B$
is determined by the qubit parameters and the external magnetic flux.

We assume, that the temperature is low enough, i.e. $k_{B}T$ is smaller than
the energy of the state $|1\rangle$, so we can neglect thermal fluctuations.

\begin{figure}
\begin{center}
\includegraphics[width=0.5\textwidth]{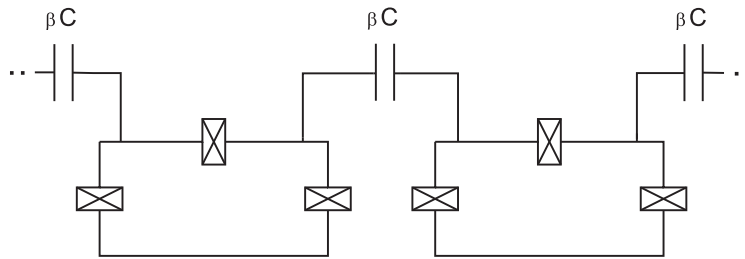}
\end{center}
\caption{Capacitively coupled qubits.}
\label{FluxChain}
\end{figure}

Persistent-current qubits can be capacitively coupled (with coupling
capacitance $\beta C$, see Fig. \ref{FluxChain}) to form a one-dimensional
array, that for $\beta \gg 1$ has the Hamiltonian

\begin{equation}
H=-\sum_{i=2}^{N}[J_{xy}(\sigma_{i}^{+}\sigma_{i-1}^{-}+
\sigma_{i}^{-}\sigma_{i-1}^{+})+J_{z}\sigma_{i}^{z}\sigma_{i-1}^{z}]-
\sum_{i=1}^{N}(\Delta\sigma_{i}^{x}+B\sigma_{i}^{z})\; .
\label{hamiltonian}
\end{equation}

The terms $J_{z}\sigma_{i}^{z}\sigma_{i+1}^{z}$ are due to the small inductive
coupling between adjacent qubits. Here $J_{z} = 2M_{q,q}I_{p}^{2}$, where
$M_{q,q}$ is their mutual magnetic inductance. The coupling constant $J_{z}$
could in principle be increased by a common Josephson junction between two
neighboring qubits \cite{levitov}. The tunneling amplitude $\Delta$ between the
states $|0\rangle$ and $|1\rangle$ of the coupled qubits differs from the value
$\Delta_{0}$ for individual non-coupled qubits, because coupling suppresses
independent tunneling events in which only one qubit changes its state. Also,
simultaneous tunneling events $|11\rangle \longleftrightarrow |00\rangle$ for
two neighboring qubits are suppressed and therefore we neglect such processes
in our model. Correlated tunneling events $|10\rangle \longleftrightarrow
|01\rangle$ are unaffected by the coupling.

The Hamiltonian (\ref{hamiltonian})
contains a term $\Delta\sum_{i}\sigma_{i}^{x}$, i.e. it
does not conserve the z-component of the total spin (which is equivalent to the
number of sites in the excited state $|1\rangle$). Therefore, the theory
proposed in \cite{bose03} is not valid in our case. However, if $\beta\gg 1$
$\Delta$ is much less than $J_{xy}$ \cite{levitov} and we can neglect this term
at first. Later we will use perturbation theory to analyze how nonzero values
of $\Delta$ affect the results.

We assume that the gate capacitances are equal to $\gamma C$. As was shown in
\cite{orlando99} using the quasiclassical approach, $\Delta_{0}$ can be
obtained as

\begin{equation}
\Delta_{0}=\sqrt{E_{J}E_{C}}\sqrt{\frac{2(4\alpha^{2}-1)}{\alpha(1+\gamma)}}
\exp(-4\sqrt{M\alpha
E_{J}}(\sqrt{1-\alpha^{2}/4}-\frac{\arccos(\alpha/2)}{2\alpha}))
\end{equation}

where

\begin{equation}
M=\frac{\hbar^{2}}{E_{C}}\frac{1+2\alpha+\gamma}{4}\; .
\label{mass}
\end{equation}

We now want to consider two interacting qubits that are coupled by a capacitor
$\beta C$, see Fig. \ref{FluxChain}. The dynamics of the qubit can be described
by the motion of a fictitious particle in a potential (see Fig.
\ref{FluxQubit}) with two local minima, that correspond to the states
$|0\rangle$ and $|1\rangle$ \cite{mooij99}.  Therefore the collective dynamics
of the two qubits can be described by the motion of a particle in a
two-dimensional potential with four minima. Without interaction the effective
mass of the particle is $M$, see Eq.~(\ref{mass}). When two qubits are
connected by the capacitor, the effective mass to move in $(0,0)\leftrightarrow
(1,1)$ direction is $M+2m^{*}$ (here $m^{*}=(\hbar/2e)^{2}\beta C$), the
effective mass for independent qubit tunneling events is $M+m^{*}$ and the
effective mass for tunneling in $(1,0)\leftrightarrow (0,1)$ direction is equal
to $M$. From these formulas one can see that the tunneling is suppressed in all
directions except $(1,0)\leftrightarrow (0,1)$, if $m^{*}\gg M$. Due to this
fact state transfer with high fidelity is possible.

Using the WKB-approach and realistic qubit parameters from
\cite{orlando99} and \cite{mooij99}, namely $\alpha=0.75$, $\gamma=0.02$, we
calculate $\Delta$ and $J_{xy}$ for our Hamiltonian:

\begin{equation}
\Delta=\Delta_{0}\exp(-0.49\sqrt{E_{J}/E_{C}}(\sqrt{1+\beta/5}-1))\; ,
\end{equation}

\begin{equation}
4J_{xy}=\Delta_{0}e^{-0.49
\sqrt{E_{J}/E_{C}}}(1-e^{-0.98\sqrt{E_{J}/E_{C}}(\sqrt{1+\beta/5}-1)})\; .
\end{equation}

With $E_{J}/E_{C}\approx 100$, we obtain

\begin{equation}
\Delta/\Delta_{0}=\exp(-4.9(\sqrt{1+\beta/5}-1))\; .
\end{equation}

Therefore, independent tunneling is effectively suppressed for $\beta\sim 10$.
$\Delta$ and $4J_{xy}$ coincide for $\beta=15$. For $\beta=20$, $4J_{xy}$ is
three times larger and for $\beta=30$ it is 25 times larger than $\Delta$. In
this case, as we will show later, $\Delta$ can be neglected.

For $\Delta=0$, the Hamiltonian (\ref{hamiltonian}) is that of an
asymmetric (XXZ) Heisenberg model in the presence of a magnetic field,

\begin{equation}
H_{L}=-\sum_{i=2}^{N}[J_{xy}(\sigma_{i}^{+}\sigma_{i-1}^{-}+
\sigma_{i}^{-}\sigma_{i-1}^{+})+J_{z}\sigma_{i}^{z}\sigma_{i-1}^{z}]-
\sum_{i=1}^{N}B\sigma_{i}^{z}\; .
\label{xxz}
\end{equation}

We now want to calculate the fidelity of the state transfer. The chain
is initialized in the state $|00...00\rangle$ by first choosing a large
negative value for the parameter $B$, see Eqs.~(\ref{xxz}) and
(\ref{bfield}).  Then, the first qubit is prepared in the state
$|\psi_{in}\rangle$ i.e., the total state of the array is
$|\psi_{in},00...00\rangle$. This is not an eigenstate of the
Hamiltonian (\ref{xxz}), therefore the system will evolve in
time. After a time $t$ the state of the last qubit is read out. In
general the last qubit will be in a mixed state, which is described by
a density matrix $\rho_{out}$. Following \cite{bose03}, we average the
fidelity over all pure input states on the Bloch sphere

\begin{equation}
F(t)=\frac{1}{4\pi}\int \langle
\psi_{in}|\rho_{out}(t)|\psi_{in}\rangle
\mathrm{d}\Omega
\end{equation}

to obtain a quantity $1/2\leq F(t)\leq 1$ that measures the quality of
transmission independent of $|\psi_{in}\rangle$.

\section{Calculation of the average fidelity}

We perform our calculations in the basis
$|k\rangle=|00...010...0\rangle$ for which the spin in the $k$-th
qubit is in the state $|1\rangle$ and all others are in the state
$|0\rangle$. The Hamiltonian (\ref{xxz}) of the array commutes with
the z-component of the total spin $\sum_{i}\sigma_{i}^{z}$. Therefore
we can use the results of \cite{bose03} to calculate the average
fidelity in terms of $f_{1,N}^{N}(t)=\langle 1|e^{-iH_{L}t}|N \rangle$
i.e., the transition amplitude of the excitation over the array.
The average fidelity can then be expressed as
\begin{equation}
F(t)=\frac{1}{2}+\frac{|f_{1,N}^{N}(t)|^2}{6}+\frac{|f_{1,N}^{N}(t)|
\cos(\gamma)}{3}\; ,
\end{equation}

where $\gamma=\mathrm{Arg}(f_{1,N}^{N}(t))$ is the argument of the
complex quantity $f_{1,N}^{N}(t)$.

Varying the magnetic field one can make $\gamma$ a multiple of $2\pi$ to
maximize the average fidelity, such that the maximum fidelity will correspond
to the maximum of $|f_{1,N}^{N}(t)|$. Furthermore, the fidelity of any state
transfer is unity, if the modulus of the amplitude to transmit the state
$|1\rangle$ across the array is unity. The fidelity for a given state
$|\psi_{in}\rangle=\cos(\theta/2)|0\rangle+
e^{i\varphi}\sin(\theta/2)|1\rangle$ in the case $f_{1,N}^{N}(t)\equiv f=|f|$
is

\begin{equation}
F(\theta,\varphi)=\frac{1+f}{2}+\cos(\theta)\frac{1-f^{2}}{2}+
\cos^{2}(\theta)\frac{f^{2}-f}{2}\; .
\end{equation}

It changes monotonically from 1 for the $|0\rangle$ state to $f^{2}$ for the
$|1\rangle$ state. For $f\neq|f|$ the fidelity can have a local minimum for
$\theta=\arccos(\frac{1-|f|^{2}}{2(|f|\cos(\gamma)-|f^{2}|)})$ if
$|f|>\cos(\gamma)/3+\sqrt{\cos^{2}(\gamma)+3}/3$.

We will now calculate $|f_{1,N}^{N}(t)|$ in the case $\Delta=0$. The
eigenfunctions of $H_{L}$ can be described as follows:

\begin{equation}
|\tilde{k}\rangle=\sum_{n=1}^{N}b_{k,n}|n\rangle\; .
\end{equation}

\begin{figure}
\begin{center}
\includegraphics[width=0.5\textwidth]{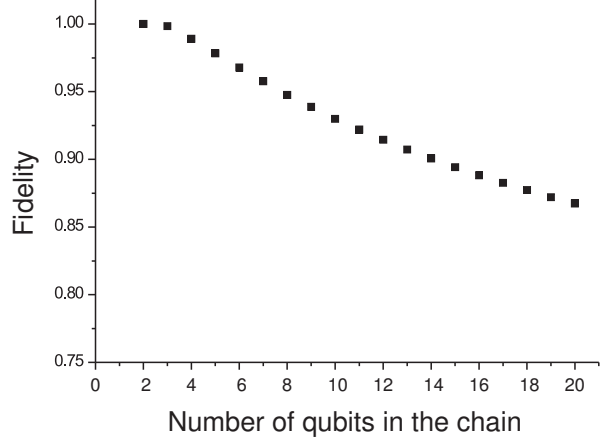}
\end{center}
\caption{First fidelity maximum for an array with $\alpha=0.75$, $\gamma=0.02$,
$E_{J}/E_{C}=100$, $\beta=30$ and $E_{J}=3$GHz, $a=0.1$.} \label{fig3}
\end{figure}

\begin{figure}
\begin{center}
\includegraphics[width=0.5\textwidth]{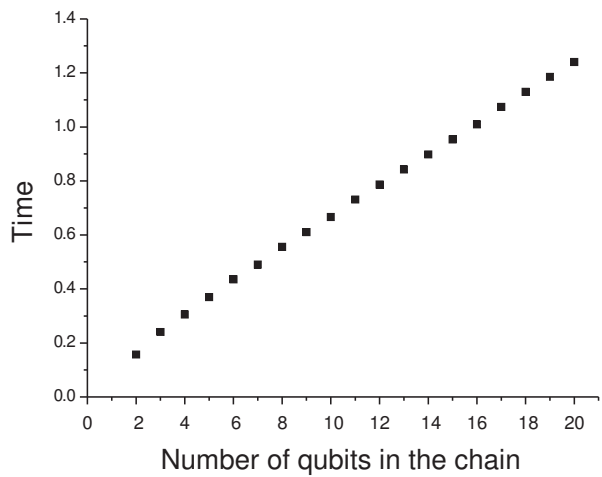}
\end{center}
\caption{Time (in units of $1/E_{J}$) at which the first fidelity maximum is
achieved. It is proportional to the length of the chain and depends on the
coefficient $J_{xy}$.}
\label{fig4}
\end{figure}

From the Schr\"odinger equation

\begin{eqnarray}
\fl H_{L}|\tilde{k}\rangle=(B(N-2)
-J_{z}(N-5))|\tilde{k}\rangle-2J_{z}(b_{k,1}|1\rangle+
b_{k,N}|N\rangle)\nonumber\\-
4J_{xy}(b_{k,2}|1\rangle+\sum_{n=2}^{N-1}(b_{k,n-1}+b_{k,n+1})|n
\rangle+b_{k,N-1}|N\rangle)\; ,
\end{eqnarray}
we obtain the following system of equations for the coefficients $b_{k,n}$

\begin{equation}
\left\{ \begin{array}{ll}
b_{k,n-1}+b_{k,n+1}=D b_{k,n} & (n\in[2,N-1])\\
a b_{k,1}+b_{k,2}=D b_{k,1} & \\
a b_{k,N}+b_{k,N-1}=D b_{k,N} &
\end{array} \right.
\end{equation}

where $a=J_{z}/2J_{xy}$ and $D$ is a constant. From the first two equations
$b_{k,i}$ can be expressed in terms of $b_{k,1}$ as

\begin{equation}
b_{k,i}=P_{i}(D_{k}) b_{k,1}\; ,
\end{equation}

here $D_{k}, k=1,...,N$ are the roots of

\begin{equation}
(D-a)P_{N}(D)=P_{N-1}(D)\; .
\end{equation}

\begin{figure}
\begin{center}
\includegraphics[width=0.5\textwidth]{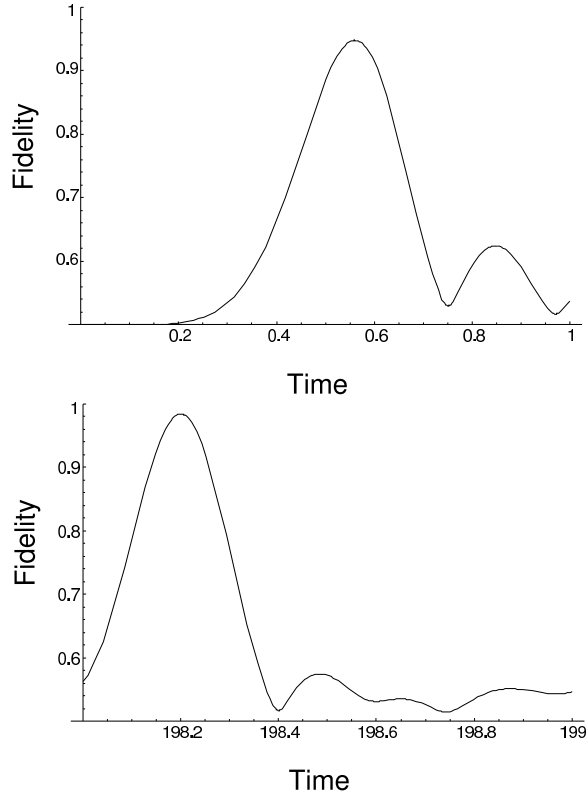}
\end{center}
\caption{Fidelity as a function of time (in units of $1/E_{J}$) for a chain
with $N=8$. Upper panel: first fidelity maximum at small times. Lower
panel: fidelity maxima around $t=198$. The parameters are chosen as in
Fig. \ref{fig3}.}
\label{fig5}
\end{figure}

$P_{i}(D)$ is a polynomial, that is determined recursively

\begin{equation}
P_{1}=1, \quad P_{2}=D-a, \quad P_{i}=DP_{i-1}-P_{i-2},\quad
i=3,...,N\; .
\end{equation}

The coefficient $b_{k,1}$ can be found from the normalization conditions

\begin{equation}
\langle\tilde{k}|\tilde{m}\rangle=\delta_{k,m} \Rightarrow
b_{k,1}^{2}=\frac{1}{P_{1}^{2}(D_{k})+...+P_{N}^{2}(D_{k})}\; .
\end{equation}

Thus we have determined the eigenfunctions of the Hamiltonian and can find its
eigenenergies

\begin{equation}
E_{k}=-J_{z}(N-5)+B(N-2)-4D_{k}J_{xy}\; .
\end{equation}

Setting $E_{0}=0$, we obtain

\begin{equation}
E_{k}=2B+4J_{z}-4D_{k}J_{xy}\; .
\end{equation}

The transition amplitude of the excitation through the array is given by

\begin{equation}
f_{1,N}^{N}(t)=\sum_{k=1}^{N}
\langle\tilde{k}|1\rangle\langle{N}|\tilde{k}\rangle e^{-iE_{k}t}
=\sum_{k=1}^{N}b_{k,1}b_{k,N} e^{-iE_{k}t}\; .
\end{equation}

%

\begin{figure}
\begin{center}
\includegraphics[width=0.5\textwidth]{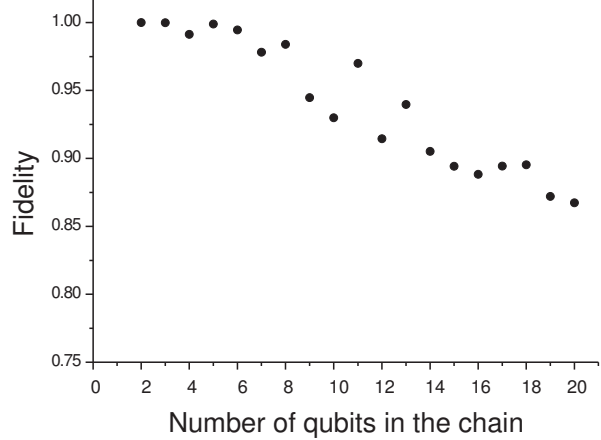}
\end{center}
\caption{Fidelity maxima for times less than $4000/J_{z}$, all the chain
parameters are as in Fig. \ref{fig3}.}
\label{fig6}
\end{figure}

\begin{figure}
\begin{center}
\includegraphics[width=0.5\textwidth]{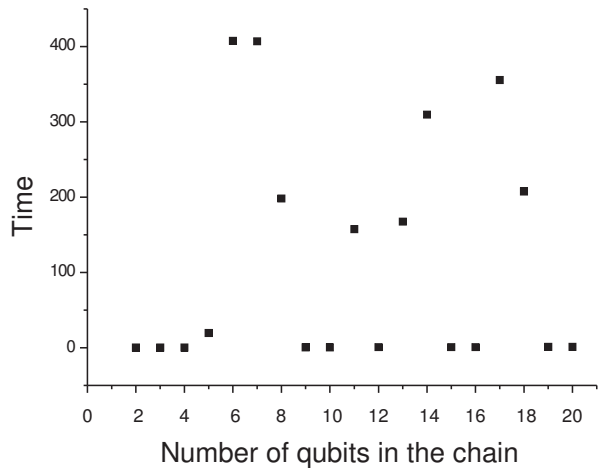}
\end{center}
\caption{Times (in units of $1/E_{J}$) at which the fidelity maxima in
Fig. \ref{fig6} are achieved.}
\label{fig7}
\end{figure}

Using these formulas we have numerically calculated the average
fidelities for different chain lengths and ratios
$a=J_{z}/2J_{xy}$. The most relevant quantities for practical purposes
are the first fidelity maxima, see Fig. \ref{fig3} and
Fig. \ref{fig4}, that we will call ``fidelity'' in the rest of this
article.

For short-length chains the average fidelity is higher than 0.9. This
makes persistent qubit arrays good candidates for transmission lines
in quantum computers, that are based on flux degrees of freedom. Also
they can be effectively used in the two-chain method proposed for
achieving perfect state transfer \cite{bose04}. The fidelity has a
complicated oscillating behavior as a function of time, see
Fig. \ref{fig5}. There are many local maxima, and the first of them is
usually not the global maximum. Therefore, waiting long enough, we can
achieve a higher fidelity. This can be seen by comparing
Fig. \ref{fig3} and Fig. \ref{fig6}.

However, the waiting times, i.e. the times, at which the maximum peaks of the
fidelity shown in Fig. \ref{fig6} occur are much longer than for the first
maximum, see Fig. \ref{fig7}. Therefore, from a practical point of view the
first maxima in the fidelity are more relevant.

Decoherence is another important reason why practical realizations of
our proposal would have to focus on the first fidelity maximum.  Like
any physical realization of a qubit, flux qubits are characterized by
a finite dephasing time, and in a recent experiment times of order
$\tau_\phi \approx 20$ns were reported for a {\it single} flux qubit
\cite{chiorescu}.  Since the time for the appearance of the first
fidelity maximum is of order $\hbar L/E_J$. As a simple estimate of
the effects of decoherence, we compare this time with the dephasing
time, which leads to a a limit of the length of the array of $L \sim
\tau_\phi E_J/\hbar \sim 100$.  Additional maxima after the first one
will be further reduced by decoherence since they correspond to states
traversing the array more than once.

To maximize the fidelity $\gamma=\mathrm{Arg}(f_{1,N}^{N}(t))$ has to
be chosen equal to zero. This can be done by varying the magnetic
field, so that $-2Bt+\gamma_{0}=2\pi n$. Here $\gamma_{0}$ is
transition amplitude phase for $B=0$. To achieve more control of the
qubit parameters the central junction can be replaced by a SQUID
\cite{orlando99}.

The works of Bose \cite{bose03} and Christandl {\it et al.} \cite{christandl}
correspond to spin chains with a particular form of the Hamiltonian $H_{L}$
($J_{z}/2J_{xy}=1$, $J_{z}/2J_{xy}=0$).
We have checked that in these limits our
results agree with \cite{bose03} and \cite{christandl}.

As mentioned above, the Hamiltonian of the real chain contains a term
$\Delta\sum_{i=1}^{N}\sigma_{i}^{x}$, that does not conserve the
z-component of the total spin (i.e. the number of
excitations). $\Delta$ is small, however, i.e.  we can use
perturbation theory to analyze the influence of this term on the
average fidelity. In this case we need to do calculations in a larger
$(2^{N}+1)$-dimensional space, because in principle any number of
excitations is possible. One can easily show, that in zero-order
approximation the fidelity and the $N+1$ lowest eigenstates will be
the same as in the unperturbed case.  The first-order corrections are
zero, because $\langle k|\sigma_{i}^{x}|k\rangle=0$ for the lowest
eigenstates. So only the second-order terms, which are proportional to
$\Delta^{2}$, affect the fidelity. The influence of the
symmetry-breaking term therefore vanishes quadratically with $\Delta$.

\begin{figure}
\begin{center}
\includegraphics[width=0.5\textwidth]{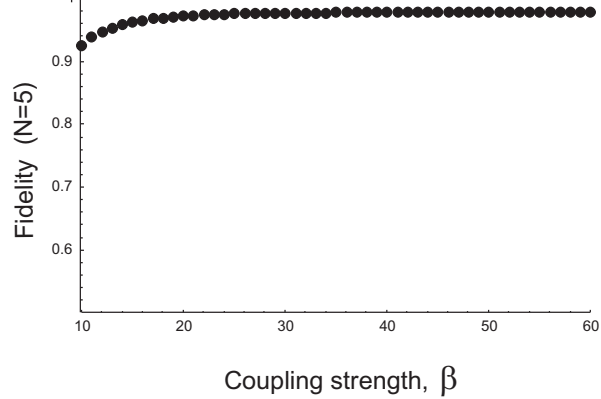}
\end{center}
\caption{Fidelity dependence on $\beta$ for chain length $N=5$.}
\label{fig8}
\end{figure}

From Fig. \ref{fig8} one can see that for qubits with the parameters
mentioned in Fig. \ref{fig3} it is sufficient to choose coupling
capacitors with about 25-30 times the junction capacitance. In this
case we can neglect the influence of $\Delta$.  For long times this
term becomes more important. This is another reason why only the first
maxima are useful for practical realizations of high-fidelity
transmission lines. One can, in principle, raise $\beta$ to make the
$\Delta$-term less important for the maxima that occur later, but in
this case the charging energy will increase and this will influence
the fidelity and the properties of the qubit.

\section{Conclusions}
We have shown that a persistent-current qubit array is a good
candidate for quantum state transfer with high fidelity in flux-qubit
based quantum computers. For short-length chains the average fidelity
of state transfer is higher than 0.9.  Therefore, this type of array
can be effectively used in the two-chain algorithm \cite{bose04} for
achieving perfect state transfer. The influence of the term
proportional to $\Delta\sigma^{x}$, that does not commute with the
z-component of the total spin, is quadratic in $\Delta$ and can be
neglected at small times for $\beta\gg 1$.

We would like to thank R. Fazio for valuable discussions.

\end{document}